\documentclass[12pt]{article}

\usepackage[body={17.5cm, 21cm},right=2cm]{geometry}

\usepackage{color}
\usepackage{graphicx}
\usepackage{epsf}
\usepackage{graphicx,epsfig}
\usepackage{amsmath}
\usepackage{amssymb}
\usepackage{epsfig}
\usepackage{cite}
\usepackage{color,colordvi}
\newcommand{\be}{\begin{eqnarray}}
\newcommand{\ee}{\end{eqnarray}}
\newcommand{\bi}{\begin{itemize}}
\newcommand{\ei}{\end{itemize}}

\newcommand{\bx}{{\vec{x}}}

\let\latexcirc=\circ
\newcommand{\ccirc}{\mathbin{\mathchoice
  {\xcirc\scriptstyle}
  {\xcirc\scriptstyle}
  {\xcirc\scriptscriptstyle}
  {\xcirc\scriptscriptstyle}
}}
\newcommand{\xcirc}[1]{\vcenter{\hbox{$#1\latexcirc$}}}
\let\circ\ccirc

\newcounter{hran}

\def\MSbar{\relax\ifmmode\overline{\rm MS}\else{$\overline{\rm MS}${ }}\fi}

\def\d{\rm d}

\def\d{{\rm d}}

\def\vq{\vec{q}}
\def\vx{\vec{x}}
\def\vv{\vec{v}}

\def\bx{{\vec{x}}}
\def\bk{{\vec{k}}}

\def\bv{{\vec{v}}}
\def\t{\tau}

\def\t{\tau}

 \def\vx{\vec{ x}}
   
\def\vk{\vec{k}}
\def\vn{\vec{n}}
\def\vq{\vec{q}}

\numberwithin{equation}{section}
\begin{document}
\thispagestyle{empty}
\vspace{5mm}
\vspace{0.5cm}
\begin{center}

\def\thefootnote{\fnsymbol{footnote}}

{\Large \bf 
Equal-time  Consistency Relations \\
\vspace{0.25cm}
in  the Large-Scale Structure of the Universe\\
\vspace{0.25cm}	
}
\vspace{2cm}
{\large  
A. Kehagias$^{a,b}$, H. Perrier$^{b}$  and A. Riotto$^{b}$
}
\\[0.5cm]

\vspace{.3cm}
{\normalsize {\it  $^{a}$ Physics Division, National Technical University of Athens, \\15780 Zografou Campus, Athens, Greece}}\\

\vspace{.3cm}
{\normalsize { \it $^{b}$ Department of Theoretical Physics and Center for Astroparticle Physics (CAP)\\ 24 quai E. Ansermet, CH-1211 Geneva 4, Switzerland}}\\

\vspace{.3cm}


\end{center}

\vspace{3cm}

\hrule \vspace{0.3cm}
{\small  \noindent \textbf{Abstract} \\[0.3cm]
\noindent 
We discuss the consistency relations involving the soft limit of the $(n + 1)$-correlator functions
 of dark matter  at equal times and their   consequences for  the halo model.
\vspace{0.5cm}  \hrule
\vskip 1cm

\def\thefootnote{\arabic{footnote}}
\setcounter{footnote}{0}


\baselineskip= 18pt

\newpage 

\section{Introduction}\pagenumbering{arabic}
Symmetry arguments  are quite useful in characterizing the  cosmological perturbations  originated  during a de Sitter  stage \cite{lrreview}.  In particular, the correlators of scalar fields, which are not the inflaton,  are constrained by conformal invariance   
\cite{antoniadis,maldacena1,creminelli1,us1,us2,vec} as the  de Sitter isometry group SO(1,4) acts  like  conformal group  on $\mathbb{R}^3$ when the fluctuations are taken on  
super-Hubble scales. Conformal consistency relations among the inflationary correlators can also be written  if the  inflationary perturbations are generated in single-field models of inflation \cite{creminelli2,hui,baumann1,baumann2,mcfadden,mata,hui2,ber}.   

Consistency relations involving the soft limit of the $(n + 1)$-correlator functions of matter and galaxy overdensities have  been recently derived by exploiting   the symmetries of the Newtonian equations of motion of the non-relativistic dark matter and galaxy  fluids coupled to gravity
\cite{KR,ppls}. These consistency relations have been  generalized to the relativistic limit \cite{jorge} (see also \cite{acc}), based on  the observation that a long mode, in single-field models of inflation, reduces to a diffeomorphism when it  is inside the horizon (as long as it is outside of the sound horizon). Furthermore, these large-scale universe consistency relations  have been generalized to
redshift space \cite{us3,cv}, their
implications for the theories of galaxy bias and modified gravity have been studied in Ref. \cite{us3} and their generalization to the case of multi-fluids has been produced in Refs. \cite{PP1,val}.

The soft limit of the $(n + 1)$-correlator functions of matter and galaxy overdensities involved in the consistency relations have the property of vanishing in the equal-time
limit. This is because they are derived by the basic property that, if an overdensity $\delta(\vx,\tau)$ (say of dark matter) satisfies the fluid equation, then also $\delta'(\vx,\tau)=\delta(\vx+\vn(\tau),\tau)$ does (here $\tau$ is the conformal time), where $\vn(\tau)$ is an arbitrary time-dependent vector. Consider the $n$-point correlation function of short modes of the density contrast. The symmetries of the Newtonian fluid equations imply that 
\be
\Big<\delta'(\vx_1)\cdots \delta'(\vx_n)\Big>=\Big<\delta(\vx_1)\cdots \delta(\vx_n)\Big>=\Big<\delta(\vx'_1)\cdots \delta(\vx'_n)\Big>.
\label{a}
\ee
If one now consider $n$-objects   contained in a sphere of radius  $R$ much smaller than the long wavelength mode of size $\sim 1/q$ and centered at the origin of the coordinates, one can choose the coordinate transformation in such a way to remove the long and linear wavelength velocity mode $\vv_L$. In such a case, 
Eq. (\ref{a})  is nothing else that the statement that the effect of a physical long wavelength velocity perturbation  onto the short modes
should be  indistinguishable from the long wavelength mode velocity generated by the transformation \cite{KR,ppls}

 \be
  \t '=\t,&~~~\bx'=\bx+\vn(\tau)=\bx-\int^\tau\d\eta\, \bv_L(\eta, \vec{0})
  =\bx+
    \frac{1}{6}\t^2\vec{\nabla}\Phi_L(\eta,\vec{0}),
 \ee
where $\Phi_L$ is the long wavelength mode of the gravitational potential. Since (${\cal H}=\dot a/a$ is the conformal Hubble rate)

\be
\Phi'(\bx,\t)&=\Phi(\bx',\t')-\left({\cal H}\dot{\vn}+\ddot{\vn}\right)\cdot \bx,
\ee
one  is basically removing the 
homogeneous gravitational force via a change of coordinates. This   corresponds to an homogeneous
acceleration transformation which allows to go to a free-falling observer, precisely the essence of the Equivalence Principle \cite{jorge}. 
In Fourier space, the transformation  (\ref{a}) gives for the short non-linear modes  

\begin{equation}
\delta'_{\vk}(\tau)=e^{i\vk\cdot\vn(\tau)}\,\delta_{\vk}(\tau)
\end{equation}
and therefore the momentum-conservation $\sum_i\vk_i=\vec{0}$ imposes the soft limit of the $(n + 1)$-correlator functions of dark matter (as well as galaxy) to vanish at equal time. 

The vanishing of the equal-time correlators in the soft limit is  therefore rooted in the fact  that one can locally eliminate the zero mode and the first spatial gradient of the  long and linear wavelength mode of the gravitational potential and that the response of the system on short scales is a uniform displacement.

In this paper, which builds up on previous works \cite{sen,zal},  we discuss what happens if we include two spatial gradients in the game, that is a non-uniform gravitational force. Based on the previous arguments, we expect that the equal-time correlators will not vanish in the soft limit, as short-scale perturbations will now feel the
non-homogeneity of the gravitational force. Indeed, we will find a consistency relation for the soft limit of the dark matter correlators which is valid
even for non-linear scales and that can be checked against analytical models modeling the clustering of dark matter on short scales.

The paper is organized as follows. In section 2 we discuss the derivation of the equal-time consistency relation, while in section 3 we check
it against the halo model. Finally, in section 4 we provide our conclusions.

\section{The equal-time consistency relation for the dark matter bispectrum}
\noindent
It is well-known that  the effect of a long wavelength mode of the gravitational potential of momentum $q$ and  including two spatial gradients on the short-scale dynamics is encodable  in a local spatial curvature $K=(3/5)(\delta_L/a)$  (valid in a matter-dominated universe and after performing an angular average),  where $\delta_L$ is the linear overdensity \cite{sen}. 
In other words, the physics in a sphere of radius $R\ll 1/q$ should not be distinguishable from the physics with the same curvature as the one induced by the linear mode. 
To account for the effects of the long wavelength mode there are three effects one should take into account. First of all, the change in  scales with respect to the case in which the local system is spatially flat. As the overdensity in the sphere is simply given by

\be
1+\delta_L=\frac{a^3}{a_K^3},
\ee
where we have indicated by $a$ the scale factor in the region outside the sphere of radius $R$ and by $a_K$ the one in the
inner region, one simply gets $a/a_K\simeq (1+\delta_L/3)$. This means that comoving momenta are shifted from $\vk$ in the spatially flat outer region to $\vk(1-\delta_L/3)$ in the inner region.

The second effect is that we have to define the short-mode  overdensities with respect to the  global average
$\overline{\rho}$ and not with respect to the local overdensity $\overline{\rho}(1+\delta_L)$. This brings an extra factor
$(1+\delta_L)$ for each density contrast on short scales. Finally, we have to account for the fact that one may trade the time variable of the
correlators with the linear growth factor $D(a)$ if no other time dependencies are present in the problem. This is true if ${\rm d}\ln D(a)/{\rm d}\ln a=\Omega^{1/2}_{\rm m}$ (being  $\Omega_{\rm m}$ the abundance of dar matter with respect to the critical one) which is the condition leading to
separability between the density contrast and the peculiar velocity at any order in standard perturbation theory
\cite{pt}. While not exactly true, the mapping between time and the linear growth factor is 
good at the ${\cal O}(10\%)$ level for most redshifts \cite{pietronirge}.
 It turns out that  in a spatially local closed universe the linear growth factor
is enhanced by a factor $(1+13/21\delta_L)$ with respect to the spatially flat case \cite{sen}. This is because 
in a spatially closed universe linear perturbations growth faster than their flat region counterparts if the curvature is positive.

All in all, the dark matter correlation functions on small scales depend on the long wavelength mode as

\begin{eqnarray}
\xi_{\delta_L}(r,a)&\simeq& \xi_{0}(r,a) \nonumber\\
&+&2\delta_L \xi_{0}(r,a)\nonumber\\
&+&\frac{1}{3}\delta_L\,r\,\frac{\partial}{\partial r}\xi_{0}(r,a)\nonumber\\
&+&\frac{13}{21}\delta_L\,\frac{\partial}{\partial\ln D(a)}\xi_{0}(r,a)+{\cal O}(\delta_L^2).
\label{aa}
\end{eqnarray}
These arguments allow to calculate the soft limit of the three-point correlators of dark matter when  the wavenumber $q$
is much smaller than the other  two: $k_1\simeq k_2\simeq 1/r\gg q$. Indeed, multiplying the result (\ref{aa}) by the long wavelength 
mode $\delta_L$, averaging over many realizations and going to momentum space, we obtain 

\be
\fbox{$\displaystyle
\Big<\delta_{\vq}(\tau)\delta_{\vk_1}(\tau)\delta_{\vk_2}(\tau)\Big>^{'\,\rm av}_{q\rightarrow 0}=P_L(q,\tau)\left[
1-\frac{1}{3}\frac{\partial}{\partial \ln k_1}+\frac{13}{21}\frac{\partial}{\partial\ln D(a)}
\right]P(k_1,\tau)$},
\label{final}
\ee
where we have used the fact that

\be
\int{\rm d}^3x \,e^{i\vk\cdot\vx}\,r\,\frac{\partial}{\partial r}\xi_{0}(r,a) =\left(-3-\frac{\partial}{\partial \ln k}\right)P(k,\tau).
\ee
The label $^{'\,\rm av}$ in the bispectrum  indicates  we have removed the momentum conservation and that  the angular  average over the angle between the long and the short wavelength modes needs to be taken:  in all our considerations we have assumed a spherical average to be able to define a constant  spatial curvature to begin with. 
As expected, the bispectrum does not vanish in the equal-time correlator. The relation (\ref{final}) extends  that one found in Ref. \cite{zal} which is valid only in the mildly non-linear regime where $P(k_1,\tau)\sim D^2(a)$. 
Of course, the relation (\ref{final}) does not hold in the
presence of a primordial non-Gaussianity as the latter introduces an extra correlation among the long and the short wavelength modes.

The generalization of Eq. (\ref{final}) to higher-order correlators is straightforward. Taking into 
account that there is a factor of $(1+\delta_L/3)$ for each position vector and a $(1+13/21 \delta_L)$ enhancement of the growth factor with respect to the spatially flat case, we get for the equal-time correlator in the long wavelength mode

\begin{eqnarray}
\Big<\delta(\vx_1,\t)\cdots\delta(\vx_n,\t)\Big>_{\delta_L}&=&\Big<\delta(\vx_1,\t)\cdots\delta(\vx_n,\t)\Big>_0\nonumber \\
&+&n\delta_L \Big<\delta(\vx_1,\t)\cdots\delta(\vx_n,\t)\Big>_0\nonumber \\
&+&\frac{1}{3}\delta_L\sum_{i=1}^n\vx_i\cdot \nabla_{\vx_i}\Big<\delta(\vx_1,\t)\cdots\delta(\vx_n,\t)\Big>_0\nonumber \\
&+&\frac{13}{21}\delta_L \frac{\partial}{\partial\ln D(a)}\Big<\delta(\vx_1,\t)\cdots\delta(\vx_n,\t)\Big>_0.
\end{eqnarray}
 Multiplying the above relation by the long wavelength mode and averaging as before, we get the relation 
 
 \be 
  \Big<\delta_{\vq}(\tau)\delta(\vec{k}_1,\tau)\cdots\delta(\vec{k}_n,\tau)\Big>^{\rm av}_{q\rightarrow 0}=P_L(q,\tau)\left[
n-\sum_{i=1}^n\left(\frac{1}{3}\frac{\partial}{\partial \ln k_i}+1\right)+\frac{13}{21}\frac{\partial}{\partial\ln D(a)}
\right]\Big<\delta(\vec{k}_1,\tau)\cdots\delta(\vec{k}_n,\tau)\Big>.
\nonumber 
\label{final-n}
\ee
Using the fact that  
\be
\sum_{i=1}^n\frac{\partial}{\partial \ln k_i}\delta (\vec{k}_t)=\vec{k}_t\nabla_{\vec{k}_t}\delta(\vec{k}_t),
\ee
where $\vec{k}_t=\vec{k}_1+\cdots \vec{k}_n$ as well as the general property  $f(x)\delta'(x)=-\delta(x)f'(x)$, we get 

\be
\Big<\delta_{\vq}(\tau)\delta(\vec{k}_1,\tau)\cdots\delta(\vec{k}_n,\tau)\Big>^{'\,\rm av}_{q\rightarrow 0}=P_L(q,\tau)\left[
1-\sum_{i=1}^n\frac{1}{3}\frac{\partial}{\partial \ln k_i}+\frac{13}{21}\frac{\partial}{\partial\ln D(a)}
\right]\Big<\delta(\vec{k}_1,\tau)\cdots\delta(\vec{k}_n,\tau)\Big>'.
\label{final-n}
\ee
Notice that all the considerations made so far are valid for the
dark matter overdensities, but not for the galaxies as the trade of the time with the linear growth factor is expected
not to hold when including  further time dependencies as galaxies form at a range of redshifts and merge.
Nevertheless, the result (\ref{aa}) should be useful to test analytical models describing the correlation functions of dark matter,  as the halo model \cite{cs1,cs2,cs}. This is what we proceed to do in the following section.

\section{The equal-time dark matter consistency relation and the halo model}
In its simplest formulation, the halo model  assumes that all matter in the universe belongs to dark matter halos, identified by their mass. Therefore, two distinct particles will either belong to the same halo or to two different ones. The power spectrum of density perturbations is   the sum of two contributions: the 2-halo term, mainly accounting for the spatial correlations of the distribution of different halos, and the 1-halo term which depends instead on the spatial distribution of matter inside a single halo. 
Clearly, while the 2-halo term is expected to describe large-scale correlations, the 1-halo term provides predictions in the nonlinear regime. 
More concretely, the  expression for the 
matter power spectrum is given by
\be
\label{eq:HaloModelPowerSpectrum}
P(k)=P_{2h}(k)+P_{1h}(k),
\ee
with the 2- and 1-halo contributions given by 
\begin{eqnarray}
\label{P2h}
P_{2h}(k,z) & = & \frac{1}{\overline{\rho}^2}\left[\prod_{i=1}^2\int {\rm d}m_i\,n(m_i,z)\,{\hat\rho}(k,m_i,z)\,\right] P_h(k,m_1,m_2),\\
\label{P1h}
P_{1h}(k,z) & = & \frac{1}{\overline{\rho}^2}\int {\rm d}m\,n(m,z)\,{\hat\rho}^2(k,m,z),
\end{eqnarray}
where  $n(m)$ is the halo mass function with $n(m){\rm d}m$ the number density of halos of mass between $m$ and $(m+{\rm d} m)$, and $\hat{\rho}(k,m,z)$ is the Fourier transform of the spatial density profile $\rho(r,m)$ of a halo of mass $m$
normalized so that $\hat{\rho}(0,m)=m$. The 2-halo term depends as well on the halo power spectrum, $P_h(k,m_1,m_2)$, describing the correlation between the centers of halos of mass $m_1$ and $m_2$. As we expect halos to be tracers of the underlying matter distribution, we can assume a linear bias relation between the halo and the matter density contrasts, so that $\delta_h \approx b_1\delta$. Thus, at large scales, the halo power spectrum can be approximated as
\be\label{eq:Ph}
P_h(k,m_1,m_2) = b_1(m_1)\,b_1(m_2)P_L(k),
\ee
where $b_1(m)$ represents the linear bias function for halos of mass $m$. Note that for Gaussian initial conditions, $b_1$ only depends on the mass $m$, as implicitly assumed in Eq. (\ref{eq:Ph}). The 2-halo term can be rewritten as
\be
P_{2h}(k,z)  =  \frac{1}{\overline{\rho}^2}\left[\prod_{i=1}^2\int {\rm d}m_i\,n(m_i,z)\,{\hat\rho}(k,m_i,z)\,b_1(m,z)\,\right] P_L(k).
\label{2h}
\ee
This description can be easily extended to the matter bispectrum. In the case of a three-point function, we should account for the possibility that the three points belong to just one, two or three dark matter halos. This means that there are now three distinct contributions to the halo model expression for the matter bispectrum, that is
\be
\label{BTOT}
\Big<\delta_{\vk_1}(\tau)\delta_{\vk_2}(\tau)\delta_{\vk_3}(\tau)\Big>\equiv B(k_1,k_2,k_3)=B_{3h}(k_1,k_2,k_3)+B_{2h}(k_1,k_2,k_3)+B_{1h}(k_1,k_2,k_3),
\ee
where
\begin{eqnarray}
\label{B3h}
B_{3h}(k_1,k_2,k_3,z) & = & {1\over\overline\rho^3}\left[\prod_{i=1}^3\int \!\!{\rm d}m_i\,n(m_i,z)\,\hat{\rho}(m_i,z,k_i)\right] B_h(k_1,m_1;k_2,m_2;k_3,m_3;z),\\
\label{B2h}
B_{2h}(k_1,k_2,k_3,z) & = & {1\over\overline\rho^3}\int\!\!{\rm d}m\,n(m,z)\,\hat{\rho}(m,z,k_1)\int\!\!{\rm d}m'\,n(m',z)\,\hat{\rho}(m',z,k_2)\,\hat{\rho}(m',z,k_3)\nonumber\\
 & & \times\, P_{h}(k_1,m,m',z)+{\rm cyc.},\\
\label{B1h}
B_{1h}(k_1,k_2,k_3,z) & = & {1\over\overline\rho^3}\int\!\ {\rm d}m\,n(m,z)\,\hat{\rho}(k_1,m,z)\,\hat{\rho}(k_2,m,z)\,\hat{\rho}(k_3,m,z).
\end{eqnarray}
In this case, while the 2-halo term depends on the halo power spectrum as in the previous case, the 3-halo term involves the halo bispectrum, $B_{h}(k_1,m_1;k_2,m_2;k_3,m_3;z)$. Assuming again a  local bias relation between halos and matter, $\delta_h(m)=f(\delta)$, expanded perturbatively as $\delta_h(m)=b_1(m)\delta+[b_2(m)/2]\delta^2 + \mathcal{O}(\delta^3)$, it is possible to derive the tree-level expression for the halo bispectrum, valid only in the large-scale limit, 
in terms of the matter power spectrum $P(k)$ and the bispectrum $B(k_1,k_2,k_3)$. This reads 
\begin{eqnarray}\label{eq:Bh}
B_h(k_1,m_1;k_2,m_2;k_3,m_3;z) & = & b_1(m_1)\,b_1(m_2)\,b_1(m_3)\,B(k_1,k_2,k_3)\nonumber\\
& & + \left[ b_1(m_1)\,b_1(m_2)\,b_2(m_3)\,P(k_1)\,P(k_2)+{\rm cyc.}  \right],
\end{eqnarray}
where $b_2(m)$ is the quadratic bias function. For Gaussian initial conditions $b_1$ and $b_2$ are scale independent. Moreover, since this equation is valid on large scales,  we can replace the matter power spectrum $P$ by its linear prediction $P_L$ and the matter bispectrum $B$ by its gravitational contribution $B_{\rm G}$

\be
\label{BG}
B_{\rm G}(k_1,k_2,k_3) = 2\,F_{2}(\bk_1,\bk_2)\,P_L(k_1)P_L(k_2) + {\rm 2~ perm.},
\ee
 $F_2$ being  the standardl kernel representing the second-order solution in perturbation theory and given by
\be
F_2(\bk_i,\bk_j)=\frac{5}{7} + \frac12\left({k_i\over k_j}+{k_j\over k_i}\right)(\hat{ k}_i\cdot\hat{k}_j) + {2\over7}(\hat{k}_i\cdot\hat{k}_j)^2.
\ee
The spatial distribution of matter in a halo of mass $m$ is specified by the halo density profile $\rho(r,m)$, interpreted as an average over all halos of the same mass. 
We  consider the Navarro, Frenk and  White  form for the halo density profile \cite{NFW} $
\rho(r)=\rho_s/\left[(r/r_s)(1+r/r_s)^2\right]$,
which assumes a universal profile as a function of $r$. The parameters $r_s$ and $\rho_s$ can be expressed in terms of the virial mass of the halo $m$ and the concentration parameter $c$. In particular, the virial mass is given by $m \equiv (4\pi/3)\,R_v^3\,\Delta_v\,\overline\rho$, with $R_v$ the virial radius, 
defined as the radius of a sphere within which the mean density of the halo is $\Delta_v$ times that of the universe. We take $\Delta_v = 200$. The  concentration parameter $c$ is defined as $c= R_v/r_s$ and is typically a function of $m$.
The halo mass function $n(m)$ \cite{zen} characterizes the number density of halos per unit mass. 
The fraction of the total mass of the universe contained in all the halos with mass in the range $m$ and $(m, m+{\rm d} m)$ can be written as
\begin{equation}
{1 \over \overline\rho}\, n(m)\, m\, {\rm d}m = f(\nu)\,{\rm d}\nu.
\end{equation}
The function $f(\nu)$ has an approximately universal form and depends on the variable
\begin{equation}
\label{nu}
\nu\equiv \frac{\delta_{\rm c}}{\sigma(m,z)},
\end{equation}
with $\delta_{\rm c}$ representing the critical density for spherical collapses (we assume the Einstein-de Sitter value $\delta_{\rm c}=1.68$ at zero redshift) while $\sigma(m,z)=D(z)\sigma(m,z=0)$ is the square root of the variance of matter fluctuations in spheres of radius $R = (3m/4\pi\overline\rho)^{1/3}$ (associated to the Fourier transform $W_R(k)$ of the top-hat function in real space), 
$\sigma^2(m)\equiv 4\pi\! \int\!{\rm d}k \,k^2\,P_L(k)\,W_{R}(k)$. We will adopt the  Sheth and  Tormen form of the halo mass function \cite{ST} expression
\be
\nu f(\nu) = A\sqrt{\frac{a\,\nu^2}{2\pi}}\left[1+\frac1{(a\,\nu^2)^{p}}\right]e^{-a\nu^2/2},
\ee
where $a=0.707$ and $p=0.3$ while $A = 0.322$ ensures a proper normalization. Finally, the bias functions can be derived from the unconditional halo mass function and, in the case of the Sheth-Tormen form, one obtains for the first two the expressions \cite{ST}
\begin{eqnarray}
\label{eq:b1G}
b_1(\nu) & = & 1+{a\,\nu^2 -1 \over \delta_{\rm c}}+{2p \over \delta_{\rm c}(1+(a\,\nu^2)^p)},\\
\label{eq:b2G}
b_2(\nu) & = & \frac{8}{21}[b_1(\nu)-1]+\frac{a\,\nu^2}{\delta_{\rm c}}\frac{a\,\nu^2-3}{\delta_{\rm c}}+
\left(\frac{1+2p}{\delta_{\rm c}}+2\frac{a\,\nu^2-1}{\delta_{\rm c}}\right)\frac{2p/\delta_{\rm c}}{1+(a\,\nu^2)^p}.
\end{eqnarray}
The requirement for the total matter density to be given by
\be
\rho(\bx)  \equiv  \overline{\rho}\left[1+\delta(\bx)\right] =
 \int \!{\rm d}m\,m\,n(m)\left[1+\sum_i \frac{b_i(m)}{i!}\delta^i(\vx)\right],
\ee
imposes the condition 
\be
\label{eq:condrho}
\int\! {\rm d}m\,m\,n(m)=\overline{\rho},
\ee
along with the constraints on the bias functions,
\be
{1\over\overline\rho}\!\!\int\!\! {\rm d}m\,m\, n(m)\, b_i(m) &=& \int\!\! {\rm d}\nu \,f(\nu)\, b_i(\nu) = \delta_{i1}.
\label{eq:condbias}
\ee
Such relations assure that, on large scales ($k \rightarrow 0$, $ \hat{\rho} \rightarrow m$), the 2-halo term of the power spectrum reduces to the linear power spectrum and the 3-halo term of the bispectrum reduces to the large-scale matter bispectrum.

We are now ready to study the squeezed limit of the bispectrum. It allows for  a significant simplification of the halo model expressions. As physical intuition dictates,   in this limit the largest contribution to the bispectrum comes from the position-space configuration where two points are close and belong to the same halo while the third one is at larger distance from the first two, and hence is likely to belong to another halo. In this case we  expect the halo model prediction to be dominated by the 2-halo contribution, with the 1-halo and 3-halo terms being subdominant. 
This expectation is confirmed numerically \cite{figueroaetal}: taking the smallest wavenumber to be 
 $q=0.014h\,{\rm Mpc}^{-1}$,   $B_{2h}$ becomes dominant over the other two terms for $k_1\sim k_2 \sim 0.3\, h\,{\rm Mpc}^{-1}$ and at 
 $k_1\sim k_2 \sim 1\, h\,{\rm Mpc}^{-1}$ the sum of  $B_{1h}$ and $B_{3h}$ contribute less than 10\%.
 
If $q$ is still in the linear regime, then we can safely set that the Fourier transform of the halo profile to be
\begin{equation}\label{eq:rho_kIR}
\hat{\rho}(m,z,q)\simeq m,
\end{equation}
in all bispectrum terms involving $q$ in Eqs.~(\ref{B3h}), (\ref{B2h}) and (\ref{B1h}). By making this substitution, the expressions for the 1-, 2- and 3-halo terms greatly simplify. Moreover, by using the conditions in Eqs. (\ref{eq:condrho}) and (\ref{eq:condbias}), together with Eq.  (\ref{eq:rho_kIR}), the halo model bispectrum contributions become, at leading order in  $q$  \cite{figueroaetal}
\begin{eqnarray}
\label{pp1}
B_{1h}(q,k_1,k_2)&= &
{1 \over \overline{\rho}}\,\epsilon_2^{[m]}(k_1),\\
\label{pp2}
B_{2h}(q,k_1,k_2)&= &
\epsilon_2^{[b_1]}(k_1)\,P_L(q),\\
\label{pp3}
B_{3h}(q,k_1,k_2)&= &
2\left[ \frac{13}{14} + \left( \frac{4}{7} - \frac12 \frac{\partial \ln P_{L}}{\partial \ln k_1}\right) (\hat q \cdot \hat{k}_1)^2 + \frac{\epsilon_{1}^{[b_2]}(k_1)}{\epsilon_{1}^{[b_1]}(k_1)}\right]\, P_L(q)\,P_{2h}(k_1).
\end{eqnarray}
The functions $\epsilon_i^{[F]}$ in these expressions
are defined as
\be
\epsilon_i^{[F]}(k) \equiv \frac{1}{\overline{\rho}^{\, i}}\int\!\! {\rm d}m\,n(m,z)\,\hat{\rho}^{\,i}(m,z,k)\,F(m,z),
\ee
where $F(m,z)$ represents a generic function of mass and redshift. Thus, these functions are like an ``average'' of the function $F$, weighted by the mass function and the $i$-th power of the Fourier transform of the density profile.
The first two terms inside the bracket of Eq.~(\ref{pp3}) have been derived by taking the squeezed limit $q \ll k_1 \simeq  k_2$ of $B_{\rm G} (k_1,k_2,k_3)$ in Eq.~\eqref{BG}
\begin{eqnarray}\label{eq:GaussSqueezed}
B_{\rm G} (q,k_1,k_2) & \simeq &2 \left[ F_2(\vq,\vk_1) P_L(q) P_L(k_1) + F_2(\vq,\vk_2) P_L(q) P_L(k_2) \right]  \nonumber\\
&=& 2\left[ \frac{13}{14} + \left( \frac{4}{7} - \frac{1}{2} \frac{\partial \ln P_{L}}{\partial \ln k_1}\right) (\hat{q}_1 \cdot \hat{k}_1)^2 + {\cal O}(q/k_1) \right] P_L(q) P_L(k_1).
\end{eqnarray}
 Let us first consider the angular averaged of the squeezed limit of the bispectrum
in the limited range of momenta where it is dominated by the 3-halo piece, that is  in the limit in which all momenta are in the linear regime, and one can safely take $P(k_1)\simeq P_{2h}(k_1)\simeq P_L(k_1)$. In such a case
\begin{eqnarray}
\Big<\delta_{\vq}(\tau)\delta_{\vk_1}(\tau)\delta_{\vk_2}(\tau)\Big>^{'\,\rm av}_{q\rightarrow 0}&\simeq& 
2\left[ \frac{13}{14} + \frac{1}{3} \cdot \frac{4}{7} - \frac{1}{6} \frac{\partial \ln P_{L}}{\partial \ln k_1}  + \frac{\epsilon_{1}^{[b_2]}(k_1)}{\epsilon_{1}^{[b_1]}(k_1)}\right]\, P_L(q)\,P_{2h}(k_1)\nonumber\\
&\simeq& 
\left[ \frac{47}{21}   - \frac{1}{3} \frac{\partial}{\partial \ln k_1}  +2 \frac{\epsilon_{1}^{[b_2]}(k_1)}{\epsilon_{1}^{[b_1]}(k_1)}\right]\, P_L(q)\,P(k_1).
\label{se}
\end{eqnarray}
\begin{figure}
\centering
\includegraphics[width=0.65\textwidth]{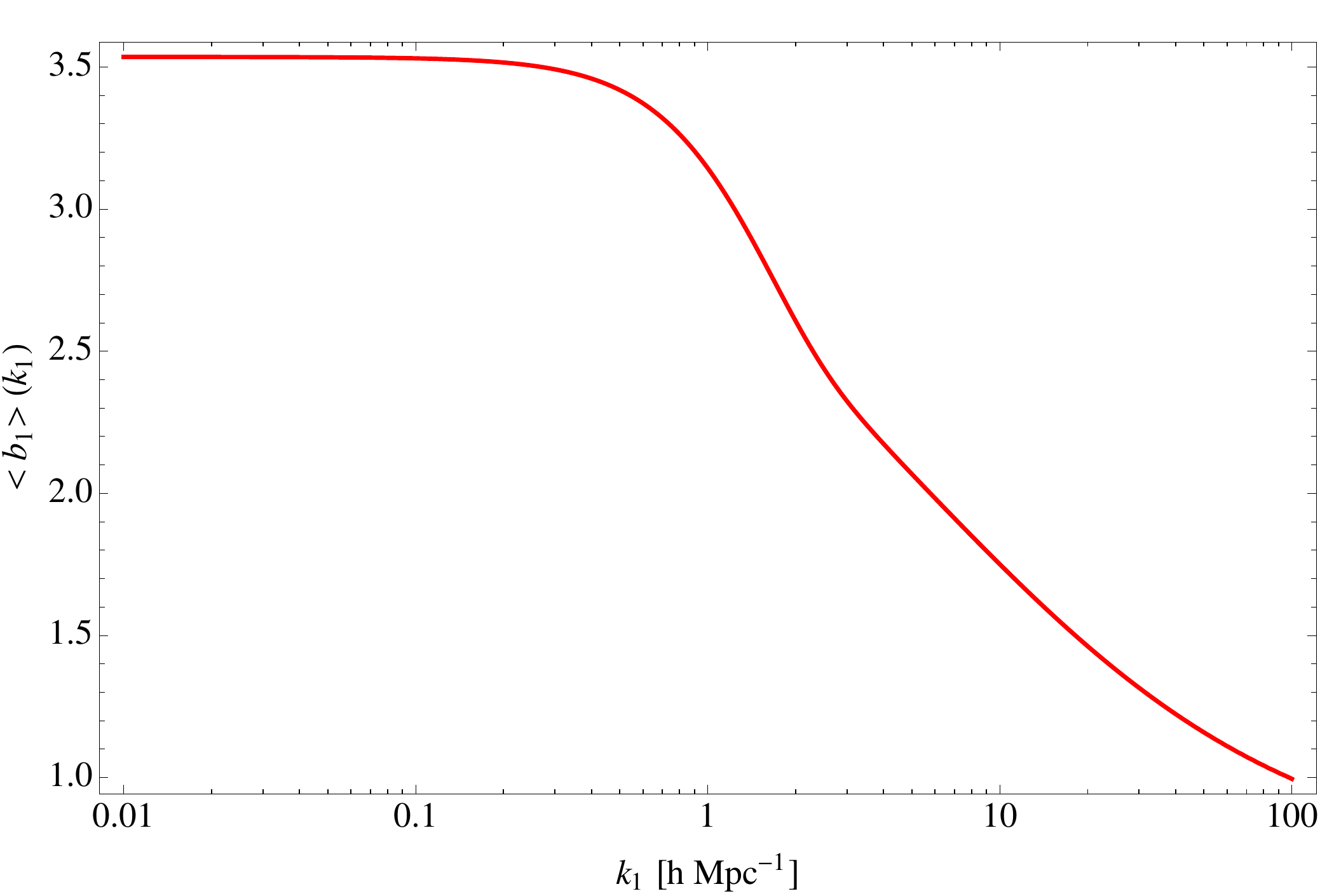}
\caption{\small The function $\langle  b_1\rangle(k_1)$ as a function of $k_1$.}
\label{bias}
\end{figure}
This has to be compared with the prediction of the consistency relation (\ref{final}) which gives, taking into account that in the present case $P(k_1)=P_L(k_1)\propto D^2(a)$,

\be
\Big<\delta_{\vq}(\tau)\delta_{\vk_1}(\tau)\delta_{\vk_2}(\tau)\Big>^{'\,\rm av}_{q\rightarrow 0}\simeq 
\left[ \frac{47}{21}   - \frac{1}{3} \frac{\partial}{\partial \ln k_1}\right]\, P_L(q)\,P(k_1).
\ee
Since  in the limit we are taking  

\begin{eqnarray}
\epsilon_{1}^{[b_2]}(k_1)&=&{1\over\overline\rho}\!\!\int\!\! {\rm d}m\,m\, n(m)\, b_2(m,k_1) =0,\nonumber\\
\epsilon_{1}^{[b_1]}(k_1)&=&{1\over\overline\rho}\!\!\int\!\! {\rm d}m\,m\, n(m)\, b_1(m,k_1) =1,
\end{eqnarray}
we conclude that the halo model prediction (\ref{se}) reproduces  the consistency relation. 

Let us now consider the most realistic case in which the momenta $k_1$ and $k_2$ are in the non-linear regime. In such a case, the bispectrum is dominated by the 2-halo piece and one gets

 \begin{eqnarray}
\Big<\delta_{\vq}(\tau)\delta_{\vk_1}(\tau)\delta_{\vk_2}(\tau)\Big>^{'\,\rm av}_{q\rightarrow 0}&\simeq& 
P_L(q)P_{1h}(k_1)\frac{\int\,{\rm d}m\,n(m,z)\,\hat{\rho}^{2}(m,z,k_1)\,b_1(k_1)}{\int\,{\rm d}m\,n(m,z)\,\hat{\rho}^{2}(m,z,k_1)}\nonumber\\
&=&\langle  b_1\rangle(k_1) P_L(q)P_{1h}(k_1),
\label{aaa}
\end{eqnarray}
where the ratio of the two integrals provides an average of the first bias parameter. 
We see from Fig. \ref{bias} that $\langle  b_1\rangle(k_1)$ is constant up to $k_1={\cal O}(1)\,h\,{\rm Mpc}^{-1}$ and its value is approximately 3.5. At much larger values of $k_1$ the average $\langle  b_1\rangle(k_1)$ acquires a strong momentum dependence.
To check if the analytical prediction of the halo models satisfies the  consistency relation we need to compute the 1-halo power spectrum. At large values of $k_1$ the power spectrum is dominated by contributions near the mass scale for which $k_1 r _s\simeq 1$ \cite{cs2}.
The  mass function behaves as ${\rm d}n/{\rm d} m\sim m^{-2}\nu^{1-2p}{\rm exp}(-a\nu^2/2)$. Since the scale radius $r_s$ depends on the mass as $r_s=R_v/c\sim m^{1/3}/m^{-(3+n)/6}=m^{(5+n)/6}$, where $n$ is the spectral index of the linear power spectrum, we find that the power spectrum at high momenta goes as \cite{cs2}

\be
P_{1h}(k_1)\sim \int {\rm d} m\, \nu^{1-2p}\hat{\rho}^2(k_1 r_s).
\ee
We change variables to $x=k_1r_s\sim k_1[m/m_*(z)]^{(5+n)/6}$, where we have defined $m_*(z)$ the mass scale for which
$\sigma\left[(m_*(z)\right]=1$. Accounting for the fact that   $\sigma(m)\sim (m/m_*)^{-(3+n)/6}$, we get 

\be
P_{1h}(k_1)\sim \left[D(a)\right]^{\frac{6}{n+3}+1-2p}k^{\gamma-3},\,\,\,\,\gamma=\frac{9+3n}{5+n}-(1-2p)\left(\frac{3+n}{5+n}\right).
\ee
The consistency relation (\ref{final}) therefore  would give

\be
\label{aaaa}
\Big<\delta_{\vq}(\tau)\delta_{\vk_1}(\tau)\delta_{\vk_2}(\tau)\Big>^{'\,\rm av}_{q\rightarrow 0}\simeq\left[
1-\frac{1}{3}(\gamma-3)+\frac{13}{21}\left(\frac{6}{n+3}+1-2p\right)
\right]P_L(q)P_{1h}(k_1),
\ee
which seems  reproduce  the result (\ref{aaa}) quite remarkably. For instance, for $n\simeq -1$, one gets a coefficient of order of 3.6  in Eq. (\ref{aaaa}) in front of $P_L(q)P_{1h}(k_1)$. On general grounds, 
as the spectral index of the short-mode power spectrum is negative and the dependence on the growth factor is such that the non-linear power spectrum is suppressed at high redshifts, implying   a positive power of $D(a)$, this leads to the conclusion that the  overall coefficient in front of $P_L(q)P_{1h}(k_1)$ in the bispectrum is predicted to be larger than unity by the consistency relation. This is in good agreement with the prediction of the halo model: 
the halo model  reproduces  the remarkable property predicted  by the consistency relation  that the bispectrum in the squeezed limit factorizes in terms of the product of the linear power spectrum times the non-linear one. This conclusion seems to be  valid up to scales ${\cal O}(1)\,h\,{\rm Mpc}^{-1}$.

\section{Conclusion}
In this paper we have derived a consistency relation which applies  for dark matter overdensities and involves the soft limit
of the $(n+1)$-point correlation functions. While this result can be hardly extended to the more interesting case of galaxy overdensities, we have used it
to investigate the ability of the halo model to satisfy the consistent relations. We have concluded that
the halo model satisfies rather well the  features predicted by the consistency relation up to scales ${\cal O}(1)\,h\,{\rm Mpc}^{-1}$.

 \section*{Acknowledgments}
 When completing this work, Ref. \cite{val2} appeared.  Our
results  agree with those contained in this reference. We thank D. Figueroa for useful numerical assistance.
  A.R. is supported by the Swiss National
Science Foundation (SNSF), project `The non-Gaussian universe" (project number: 200021140236). 
The  research of A.K. was implemented under the ``Aristeia" Action of the 
``Operational Programme Education and Lifelong Learning''
and is co-funded by the European 
Social Fund (ESF) and National Resources.  It is partially
supported by European Union's Seventh Framework Programme (FP7/2007-2013) under REA
grant agreement n. 329083.


\end{document}